\documentstyle[aps,prl,epsf]{revtex}

\begin{document}
\twocolumn[\hsize\textwidth\columnwidth\hsize\csname
@twocolumnfalse\endcsname
\title{$d_{x^2-y^2}$ superconductivity in a generalized Hubbard model.}
\author{Liliana Arrachea$^{a}$ and A.A. Aligia$^{b}$.}
\address{$^{a}$PUC-Rio, Departamento de Fisica,\\
Caixa Postal 38071, Cep: 22452-970 - RJ, Rio de Janeiro, Brasil\\
$^{b}$Centro At\'{o}mico Bariloche and Instituto Balseiro,\\
Comisi\'on Nacional de Energ\'{\i}a At\'{o}mica, \\
8400 Bariloche, Argentina.}
\maketitle

\begin{abstract}
We consider an extended Hubbard model with nearest-neighbor correlated
hopping and next nearest-neighbor hopping $t^{\prime }$ obtained as an
effective model for cuprate superconductors. Using a generalized
Hartree-Fock BCS approximation, we find that for high enough $t^{\prime }$
and doping, antiferromagnetism is destroyed and the system exhibits $d$-wave
superconductivity. Near optimal doping we consider the effect of
antiferromagnetic spin fluctuations on the normal self-energy using a
phenomenological susceptibility. The resulting superconducting critical
temperature as a function of doping is in good agreement with experiment.
\end{abstract}


\vskip2pc]

\narrowtext

\section{Introduction}

One of the most challenging properties that any effective one-band model for
the high superconducting critical temperature (high-$T_{c}$) superconducting
materials should explain is the experimentally observed dependence of $T_{c}$
with doping ($x$). Extended $t-J$ models seem to be successful to interpret
several important properties of these materials \cite{an,t-J1,pla}. They
were derived as effective Hamiltonians in the strong-coupling limit (very
large $U$)\cite{zha,tpri,fei,bel}, while in generalized Hubbard models
obtained by similar derivations, $U$ is of the order of the unperturbed band
width $8t$ $\sim 3$eV \cite{fed,sim}. While numerical studies in $t-J$-like
models unambiguously indicate that they exhibit sizable superconducting
correlations with $d$-wave symmetry, in agreement with experiments \cite
{t-J1}, this is not the case of the usual Hubbard model \cite{hub}. Apart
from quantitative reasons concerning the strength of the local Coulomb
repulsion, it is of interest to study an accurate enough effective model
amenable to weak coupling many-body treatments. In particular, the observed
pseudogap behavior for temperatures $T_{c}<T<T^{*}$ in angle-resolved
photoemission spectroscopy (ARPES) experiments \cite{arpse} has been
interpreted as a precursor effect of the antiferromagnetic (AF) \cite{precaf}
as well as the superconducting state \cite{rande,eng} between other possible
scenarios \cite{em,other}. For the attractive (negative $U$) Hubbard model, a
good deal of research helped to elucidate the nature of the superconducting
transition \cite{negu} and most of this work is based on one appealing
feature of this model: its explicit attractive interaction leading to a
superconducting state in the Bardeen-Cooper-Schrieffer (BCS) approximation.
Thus, even though the symmetry of the BCS gap is $s$-wave, this model has
been used for a qualitative understanding of ARPES experiments \cite{rande}.
Due to the lack of a genuine microscopic $d$-wave superconducting analogous
of the of the negative-$U$ Hubbard model, some phenomenological models with
BCS-like interactions were studied \cite{eng,feh,avh}. While the model
proposed in Ref. \cite{feh} is built up by pure phenomenological
interactions with several free parameters, in order to fit the
experimentally observed $T_{c}$ vs $x$ curve, the AF and van Hove picture 
\cite{avh} has been proposed on the basis of a simplification of the $t-J$
model in which holes are constrained to move only in one sublattice (up or
down) in a Neel spin background. Nevertheless, this is an oversimplification
for realistic $J<t$ \cite{com}. The assumption of pairing coexisting with
long-range AF order is shared with the strong-coupling polaron picture \cite
{pla}, which also reproduces qualitatively the observed $T_{c}$ vs $x$
curve. However, nuclear magnetic resonance experiments indicate that the
coherence length of the AF correlations $\xi $ is of only a few lattice
sites in the optimally doped materials \cite{susc}. An effective attractive
separable potential has also been considered to study the $d$-wave
superconducting pseudogap evolution \cite{eng}. In this case, however, the
AF correlations are not taken into account at all, while they play a
relevant role for underdoped and optimally doped materials \cite{precaf,chu}.

In this work, we calculate $T_{c}$ vs $x$ of an effective one-band model for
the high $T_{c}$ cuprates using weak-coupling techniques. We obtain beyond a
certain doping, a stable paramagnetic phase with $d$-wave superconductivity
(DWS). The mechanism for superconductivity is the electron-hole symmetric
correlated hopping studied previously in one and two dimensions (2D) \cite
{tabmt,physc}. However, previously in 2D, the next-nearest-neighbor hopping $%
t^{\prime }$ was neglected and only $s$-wave superconductivity was found for
doping high enough to inhibit long-range antiferromagnetism. In simplified
terms, this can be understood as follows: on-site Coulomb repulsion $U$
inhibits on-site pairs, but nearest-neighbor singlet pairs are favored by
the correlated hopping. In mean field, the dependence of this effective
attraction in reciprocal space is proportional to $(\cos k_{x}-\cos
k_{y})^{2}$ for $d$ symmetry and to $(\cos k_{x}+\cos k_{y})^{2}$ for
extended $s$ symmetry (see Section III or Ref. \cite{physc}). This implies
that $d$-wave superconductivity is favored with respect to the $s$-wave one,
for doping $x$ at which the points $X$ ($(0,\pi )$ and equivalent) lye near
the Fermi surface. However, when $t^{\prime }=0$, this happens for $x\simeq
0 $ and for these dopings, due to the perfect nesting of the Fermi surface,
the antiferromagnetic instability is the dominant one and inhibits
superconductivity. When $t^{\prime }$ is included, the perfect nesting is
destroyed and the $X$ points lye at the Fermi surface for finite doping,
stabilizing the $d$-wave solution.

In Section II we briefly describe the derivation of the effective model and
explain the meaning of the different terms. In Section III we calculate $%
T_{c}$ vs $x$ using a generalized Hartree-Fock (HF) BCS decoupling. Since
the above mentioned decoupling exaggerates the range of stability of the AF
phase, in Section IV we include the effect of the AF spin fluctuations
(AFSF) for dopings at which long-range AF order is destroyed, by evaluating
the normal self-energy in the one-loop approximation with the
phenomenological susceptibility of Ref. \cite{susc}. Section V contains a
discussion.

\section{ The effective model}

The model used is obtained from the mapping of the low-energy part of the
Hilbert space of the three-band Hubbard model $H_{3b}$\cite{fed,sim}. As
first suggested by Zhang and Rice \cite{zha}, and later confirmed by
analytical and numerical work of several groups \cite{tpri,fei,bel}, when
insulating cuprates are doped, the added holes form local singlets involving
a Cu 3d$_{x^{2}-y^{2}}$ hole at a given site $i$ (with creation operator $%
d_{i\sigma }^{\dagger }$) and a hole in the linear combination $\alpha
_{i\sigma }^{\dagger }$ of the four nearest-neighbor oxygen 2p$_{\sigma }$
orbitals with $x^{2}-y^{2}$ symmetry. This Zhang-Rice singlet has the form $%
[A(\alpha _{i\uparrow }^{\dagger }d_{i\downarrow }^{\dagger }-\alpha
_{i\uparrow }^{\dagger }d_{i\downarrow }^{\dagger })+Bd_{i\uparrow
}^{\dagger }d_{i\downarrow }^{\dagger }+C\alpha _{i\uparrow }^{\dagger
}\alpha _{i\downarrow }^{\dagger }]|0\rangle $ and is mapped into the vacuum 
$|0\rangle $ at site $i$ in the effective one-band model $H$. Similarly, the
ground state of $H_{3b}$ for one hole at cell $i$, (which has the form $%
(Dd_{i\sigma }^{\dagger }+E\alpha _{i\sigma }^{\dagger })|0\rangle $) is
mapped onto $c_{i-\sigma }^{\dagger }|0\rangle $ of $H$, where $c_{i\sigma
}^{\dagger }$ is an effective electron creation operator. The vacuum of $%
H_{3b}$ at cell $i$ (which corresponds to full 3d and 2p shells) is mapped
onto $c_{i\uparrow }^{\dagger }c_{i\downarrow }^{\dagger }|0\rangle $ in $H$%
. Calculating the matrix elements in the reduced Hilbert space, and
retaining only the most important terms, the following effective Hamiltonian
results: 
\begin{eqnarray}
H &=&U\sum_{i}n_{i\uparrow }n_{i\downarrow }\;-\;\mu \sum_{i}(n_{i\uparrow
}+n_{i\downarrow })  \nonumber \\
&-&\sum_{<ij>\sigma }(c_{i\bar{\sigma}}^{\dagger }c_{j\bar{\sigma}%
}+h.c)\{t_{AA}\;(1-n_{i\sigma })(1-n_{j\sigma }) \nonumber\\
&+&t_{BB}\;n_{i\sigma}n_{j\sigma }  
+t_{AB}\;[n_{i\sigma }(1-n_{j\sigma })\nonumber\\
&+&n_{j\sigma }(1-n_{i\sigma
})]\}-t^{\prime }\sum_{<ij^{\prime }>\sigma }\;c_{i\sigma }^{\dagger
}c_{j^{\prime }\sigma },  \label{1}
\end{eqnarray}
where $<ij>$ ($<ij^{^{\prime }}>$) denotes nearest-neighbor
(next-nearest-neighbor) positions of the lattice. Note that $U$ is not
directly related with a Coulomb repulsion, but represents the cost in energy
of constructing a Zhang-Rice singlet from two singly occupied cells. It is
lower than the difference between $p$ and $d$ on-site energies of $H_{3b}$
(usually called charge-transfer energy $\Delta $). $t_{AA}$ represents the
hopping of a Zhang-Rice singlet to a singly occupied nearest-neighbor cell.
The terms with amplitude $t_{AB}$ correspond to destruction of a Zhang-Rice
singlet and a nearest-neighbor cell without holes, creating two singly
occupied cells and vice versa. $t_{BB}$ describes the movement of an
isolated hole in $H_{3b}$. Clearly, the amplitude of these three correlated
hopping processes should differ in general. The dependence of the
next-nearest-neighbor hopping on the occupation of the sites involved in the
hopping is neglected. Note that in the simplest strong-coupling derivation
leading to the $t-J$ model \cite{zha}, $t_{AA}$ amounts to the only kinetic
term in the model, the one with $t_{BB}$ is mapped out of the relevant
Hilbert-space while the one with $t_{AB}$ is treated in second order of
perturbation theory to define the exchange term \cite{fei}. More systematic
strong-coupling derivations contain other terms as well as higher-order
corrections \cite{tpri,fei,bel,sim}.

For $t^{\prime }=0$, the occurrence of metal-insulator transitions and
superconductivity have been investigated for particular cases of this model 
\cite{tabmt,physc}, but a stable DWS phase has not been found before. We
concentrate on the parameter regime $t_{AB}>$ $t_{AA}=t_{BB}=t$, which seems
more adequate for the cuprates \cite{fed,sim,physc}. The effect of $%
t^{\prime }$ \cite{tpri} is crucial to stabilize the $d$-wave phase, because
it removes the perfect nesting at half-filling and shifts the energy of the
van Hove singularity (VHS) in the unperturbed density of states (originated
by the saddle points at $X$ in the dispersion relation) away from the Fermi
energy at half filling.

\section{Mean-field approach.}

The correlated hopping terms of the Hamiltonian (\ref{1}) can be separated
in one- two- and three-body contributions, with coefficients $t=t_{AA}$, $%
t_{2}=t_{AA}-t_{AB}$ and $t_{3}=2t_{AB}-t_{AA}-t_{BB}$ respectively \cite
{physc}. We treat Eq. (\ref{1}) within the generalized HF BCS approximation 
\cite{physc}. The term in $t_{3}$ contributes to the BCS solution in the $d$%
-wave as well as in the $s$-wave channels. The self-consistent parameters
considered in the decoupling are $\langle n_{i\uparrow }\rangle -\langle
n_{i\downarrow }\rangle =me^{i{\bf Q\cdot R}_{i}}$, $\tau =\langle c_{i+{\bf %
\delta }\sigma }^{\dagger }c_{i\sigma }\rangle $, $\psi =\langle
c_{i\uparrow }^{\dagger }c_{i\downarrow }^{\dagger }\rangle $ and $\varphi
_{\delta }=\langle c_{i+{\bf \delta }\uparrow }^{\dagger }c_{i\downarrow
}^{\dagger }\rangle $, with ${\bf \delta }={\bf x,y}$ being vectors
connecting nearest neighbors and $\varphi _{x}=\pm \varphi _{y}$ \cite{physc}%
. $n=n_{\uparrow }+n_{\downarrow }=1-x$ is the particle density, $L$ is the
number of lattice-sites, ${\bf Q}=(\pi ,\pi )$ and ${\bf R}_{i}$ indicates
the lattice-position. The problem is reduced to a one-particle one with
three possibilities for the symmetry breaking perturbation: (i) AF spin
density wave (SDW) (with $m\neq 0$ and $\psi =\varphi _{x}=0$),{\em \ } (ii)
DWS (with $m=\psi =0$ and $\varphi _{x}=-\varphi _{y}\neq 0$), and (iii)
extended $s$-wave superconductivity (SWS) (with $m=0$, $\psi \neq 0$ and $%
\varphi _{x}=\varphi _{y}\neq 0$). The possibility of coexistence of SDW and
superconductivity is left out here, since a previous study indicated that a
sizeable superconducting gap is not possible within our model in presence of
long-range antiferromagnetism \cite{afd}. For the three cases, the
renormalized dispersion relation, effective hopping and effective chemical
potential can be written in the form:

\begin{eqnarray}
\epsilon _{k}\; &=&\;-2t_{eff}(\cos k_{x}+\cos k_{y})-4t^{\prime }\cos
k_{x}\cos k_{y},  \nonumber \\
t_{eff} &=&t-t_{2}n+t_{3}[3\tau ^{2}+\psi ^{2}+\varphi
_{x}^{2}-(n^{2}-m^{2})/4]  \nonumber \\
\mu _{eff}\; &=&\;\mu -(Un/2+8t_{2}\tau +4t_{3}[\tau n+ \nonumber \\
& &\psi (\varphi
_{x}+\varphi _{y})]).
\end{eqnarray}
The SDW, $d$-wave, and $s$-wave BCS order parameters are: 
\begin{eqnarray}
\Delta ^{SDW}\; &=&\;(\frac{U}{2}+4t_{3}\tau )\;m  \nonumber \\
\;\Delta _{Dk}^{BCS}\; &=&\;4t_{3}\tau \varphi _{x}(\cos k_{x}-\cos k_{y}), 
\nonumber \\
\Delta _{Sk}^{BCS}\; &=&(8t_{3}\tau -U)\psi \;-4(2t_{2}+nt_{3})\varphi _{x} 
\nonumber \\
&&+[4t_{3}\tau \varphi _{x}-2(2t_{2}+nt_{3})\psi ] \nonumber\\
& &(\cos k_{x}+\cos k_{y}),
\label{4}
\end{eqnarray}
respectively. For a given wave vector {\bf k}, they coincide with half of
the corresponding energy gap. For both, $d$- and $s$-wave superconducting
solutions, the dependence of $T_{c}$ vs $x$ is obtained from the linearized
BCS gap equations. For the $d$-wave case, $T_{c}$ is given by: 
\begin{equation}
1\;=\;\int \frac{d^{2}k}{(2\pi )^{2}}\;\tanh (\frac{{\bar{\epsilon}}_{k}}{2T}%
)\;\frac{4t_{3}\tau (\cos k_{x}-\cos k_{y})}{2{\bar{\epsilon}}_{k}}\;\cos
k_{x},  \label{5}
\end{equation}
where ${\bar{\epsilon}}_{k}=\epsilon _{k}-\mu _{eff}$, and $T$ is the
temperature. For the $s$-wave solution, $T_{c}$ is the temperature at which $%
\lambda _{max}$ equals one, being $\lambda _{max}$ the largest eigenvalue of
the matrix 

\begin{equation}
\left( 
\begin{array}{ll}
\int \frac{d^{2}k}{(2\pi )^{2}}\tanh (\frac{{\bar{\epsilon}}_{k}}{2T})\frac{%
\alpha _{k}}{2{\bar{\epsilon}}_{k}} & \int \frac{d^{2}k}{(2\pi )^{2}}\tanh (%
\frac{{\bar{\epsilon}}_{k}}{2T})\frac{\beta _{k}}{2{\bar{\epsilon}}_{k}} \\ 
\int \frac{d^{2}k}{(2\pi )^{2}}\tanh (\frac{{\bar{\epsilon}}_{k}}{2T})\frac{%
\alpha _{k}\cos k_{x}}{2{\bar{\epsilon}}_{k}}\; & \int \frac{d^{2}k}{(2\pi
)^{2}}\tanh (\frac{{\bar{\epsilon}}_{k}}{2T})\frac{\beta _{k}\cos k_{x}}{2{%
\bar{\epsilon}}_{k}}
\end{array}
\right) ,  \label{s5}
\end{equation}
where 
\begin{eqnarray}
\alpha _{k}\; &=&\;8t_{3}\tau -U-2(2t_{2}+nt_{3})(\cos k_{x}+\cos
k_{y}),\;\;\;\;  \nonumber \\
\;\beta _{k}\; &=&\;-4(2t_{2}+nt_{3})+4t_{3}\tau (\cos k_{x}+\cos k_{y}),
\label{s2}
\end{eqnarray}
are the coefficients of $\psi $ and $\varphi _{x}$ in $\Delta _{Sk}^{BCS}$.
This method of obtaining $T_{c}$ in second-order phase transitions when the
thermodynamic potential $\Omega $ depends on more than one parameter (here $%
\psi $ and $\varphi _{x}$), has been used before \cite{tra}, and is
equivalent to the usual one of finding the first instability (as $T$ is
lowered) of the Hessian matrix formed by the second derivatives of $\Omega $
with respect to the independent variables \cite{kik}.

\begin{figure}
\narrowtext
\epsfxsize=3.3truein
\vbox{\hskip 0.05truein \epsffile{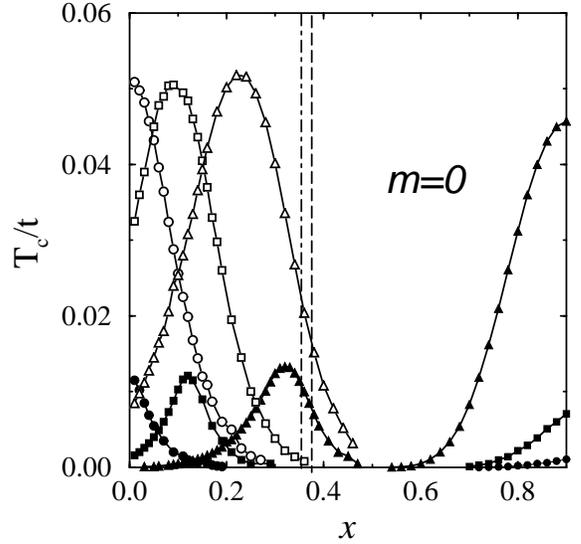}}
\medskip
\caption{$d$-wave BCS- critical temperature $T_{c}$ in units of $t$ as
a function of the doping $x$ (for $x<0.5$), for $t_{AB}=2t$ (open symbols)
and $t_{AB}=1.5t$ (filled symbols). Circles, squares and triangles
correspond to $t^{\prime }=0,\;-0.2t,\;-0.45t$, respectively. The solid
symbols for $x>0.5$ show $T_{c}$ of the $s$-wave solution for $t_{AB}=1.5$
and the same values of $t^{\prime }$ as those considered for the $d$-wave.
The dot (dot-dashed) line indicates the boundary of the SDW at $T=0$ for $%
U=4t\;$ ($U=6t$), $t_{AB}=1.5t\;$ ($t_{AB}=2t$) and $t^{\prime }=-0.45t$.}
\end{figure}

In Fig. 1, we show the superconducting critical temperature as a function of
doping $x=1-n$, where $n$ is the number of electrons per site for both, $d$%
-wave and $s$-wave solutions. Most of the pairing terms, as well as the
additional contribution to the usual $U/2$ term in the expression of the SDW
gap, can be generated from the mean-field decoupling of an effective
two-body interaction of the form \cite{afd} 
\begin{equation}
V_{eff}\;=\;t_{3}\tau \sum_{<ij>}(4{\bf S}_{i}{\bf \cdot S}_{j}+n_{i}n_{j}).
\label{vef}
\end{equation}
Paring in both $d$-wave and extended $s$-wave channels, is originated by the
spin-flip terms of $V_{eff}$. For the $s$-wave case, there are some
additional contributions, as can be observed in (\ref{4}). From mere
inspection of the three possible order parameters (\ref{4}), it can be seen
that $\Delta ^{SDW}$ and $\Delta _{Sk}^{BCS}$ depend on $U$ in such a way
that the antiferromagnetic ($s$-wave-BCS) solution is enhanced (weakened) as 
$U$ increases, whereas $\Delta _{Dk}^{BCS}$ does not depend on $U$. Thus,
the doping region where the $d$-wave BCS 
solution exists is limited only by
the difference in energy with the two other competing instabilities of the
Fermi liquid. As mentioned in Section I, due to the wave-vector dependence
of the superconducting order parameters, the shape of the Fermi-surface
plays an important role in defining the doping regions at which each of the
two BCS solutions are possible. In Ref. \cite{physc} we have shown that for $%
t^{\prime }=0$ and assuming a constant density of states, SWS exists for
high enough $x$ and low enough $U$, while DWS exists near half-filling.
Within this region, however, the SDW solution has lower energy than the DWS
one, being the difference between the energy of both solutions small for $%
U\sim 0$ (as can be inferred from the form of the gaps, Eqs (\ref{4})). When 
$t^{\prime }$ is turned on, the VHS is displaced, lying at the Fermi energy
for a finite doping. The increment in the density of states at finite doping
enhances the critical temperature as well as the doping range associated to
the $s$-wave solution. However, as a consequence of the vector dependence of
the effective interaction, the optimal doping for SWS is always high. In
Fig. 1, it is shown that for $t_{AB}=1.5$ and $U=4t$, the doping region with
SWS is $x>0.6$, i.e. beyond the range of doping accessible experimentally in
the cuprates. For the $d$-wave solution, the maximum of $T_{c}$ occurs when
the chemical potential coincides with the energy of the VHS. Due to the loss
of perfect nesting and the closing of the indirect SDW gap for this doping,
the SDW is weakened. Thus, even for large $U$, for sizeable $t^{\prime }$
there exists a doping range for which the $d$-wave BCS solution is stable
and robust. We indicate in the figure the doping region for the existence of
the SDW solution ($m\neq 0$) at $T=0$ for the particular values $U=4t\;$ ($%
U=6t$), $t_{AB}=1.5t\;$ ($t_{AB}=2t$) and $t^{\prime }=-0.45t$. The Neel
temperature at which $m$ becomes zero is not calculated here since: it is
determined by spin fluctuations, and the mean-field value is too large. For
nearly realistic bare parameters, the doping range for an stable DWS (beyond
the dot-dashed line in Fig. 1) corresponds to the overdoped regime of the
cuprates, in agreement with experiments on Bi2212 \cite{arpse} which
indicate that the superconducting state takes place in an otherwise ordinary
Fermi liquid for overdoped compounds.

In the optimally doped regime, spin fluctuations not only affect directly
the SDW order parameter, destroying the AF long-range order, but also affect
indirectly the superconducting order parameters. Their effect should be
included in a more realistic calculation of the superconducting critical
temperature. This is the goal of the next section.

\section{The effect of spin fluctuations}

Long-range antiferromagnetism is not expected to survive within the whole
range of doping predicted by the mean-field approach. For doping larger than 
$x\sim 0.05-0.1$, the spin fluctuations are important and a Fermi liquid
with strong AFSF is a more appropriate picture for this regime \cite
{susc,mon}. AFSF of any Hubbard-like microscopic model as the one considered
here can be characterized by a spin susceptibility of the form 
\begin{equation}
\chi ({\bf q},\omega )\;=\;\frac{\chi ({\bf Q},0)}{1+({\bf q}-{\bf Q}%
)^{2}\xi ^{2}-i\omega /\omega _{sf}},  \label{6}
\end{equation}
where the parameters scale with the AF correlation length $\xi $ as $\chi (%
{\bf Q},0)\sim \xi ^{2}$, $\omega _{sf}=\Gamma _{0}\xi ^{2}$, with $\Gamma
_{0}\sim 40$meV \cite{susc,vil}. Assuming a phenomenological spin-fermion
coupling, AFSF has been suggested to mediate the pairing in the cuprates 
\cite{mon}. This picture is supported by the observed correlation between $%
\xi $ and $T_{c}$ in several superconducting cuprates. Within the
weak-coupling formalism, DWS due to correlated hopping is inhibited in a
background with long-range AF correlations \cite{afd}. However, the
effective interaction Eq. (\ref{vef}) provides an explicit channel for the
coupling with collective AFSF within the doping region without AF long-range
order, which has the same form as the phenomenological coupling used in Ref. 
\cite{mon}. Thus, it might be expected that the AFSF would renormalize the
bare value of $t_{AB}$ ( $<1.5t$ for realistic values of the three-band
parameters \cite{sim} ) to higher ones \cite{chu,mon}. In what follows, we
investigate how the BCS- $T_{c}$ vs $x$ dependence is modified by the effect
of the AFSF for the $d$-wave solution.

For finite $t^{\prime }$, the Fermi surface contains hot spots (for which $%
\epsilon _{k}=\epsilon _{k+Q}$). Fermions located in the neighborhood of
these points are the most affected by AF correlations and exhibit a peculiar 
$T$-dependence in the one-particle spectral properties, which is mainly
determined by the magnitude of $\xi /\xi _{th}$, and $\omega _{sf}/T$ \cite
{chu,vil}, with $\xi _{th}=v_{F}/T$, being $v_{F}$ the Fermi velocity. Hot
spots are located at ${\bf k}_{hs}$ near $X$ ($(0,\pi )$ and symmetry
related points), i.e. near the antinodes of the DWS gap and with energies
close to the VHS (for which $v_{F}\sim 0$). As $T_{c}$ vs $x$ predicted by
the BCS approximation, as well as the value of the maximum $T_{c}$ itself,
depend on the behavior of the density of states, AFSF are expected to play
some further important role apart from the eventual renormalization of the
effective pairing interaction. The self-energy obtained from (\ref{6}) is 
\begin{eqnarray}
\Sigma ({\bf k},i\omega _{n}) &=&T\sum_{m}\int \frac{d^{2}q}{(2\pi )^{2}}g(%
{\bf q}){\bar{g}({\bf q})} \nonumber\\
& &\chi ({\bf q},i\nu _{m})G^{0}({\bf k}+{\bf q}%
,i\omega _{n}+i\nu _{m}),  \nonumber \\
\chi (q,i\nu _{m}) &=&-\int_{-\omega _{0}}^{\omega _{0}}\frac{d\omega }{\pi }%
\frac{\text{Im}\chi ({\bf q},\omega )}{i\nu _{m}-\omega },  \label{7}
\end{eqnarray}
where $[G^{0}({\bf k},i\omega _{n})]^{-1}=i\omega _{n}-{\bar{\epsilon}}_{k}$%
, $g({\bf q})$ is an effective interaction between fermions and spin
fluctuations, $\omega _{0}$ is a frequency cut off, $\nu _{m}=2m\pi T$ and $%
\omega _{n}=(2n+1)\pi T$. As usual, an effective coupling constant $%
g^{\prime }\sim g({\bf q}){\bar{g}({\bf q})}\chi ({\bf Q},0)/\xi ^{2}$ is
defined, which in the present case should be proportional to ($U+4t_{3}\tau $
) \cite{afd}. The ensuing spectral function is $A({\bf k},\omega )=-$Im$G(%
{\bf k},\omega )/\pi $, with $[G({\bf k},\omega )]^{-1}=\omega -{\bar{%
\epsilon}}_{k}-\Sigma ({\bf k},\omega )$. To examine how the changes in the
behavior of $A({\bf k},\omega )$ affect the $T_{c}$ vs $x$ dependence, we
consider the effect of AFSF using the BCS form of the anomalous self energy $%
\Delta _{Dk}^{BCS}$ given by Eq. (\ref{4}) and calculating the normal self
energy in the one-loop approximation ( Eq. (\ref{7})). The resulting
linearized gap equation, 
\begin{eqnarray}
\varphi _{x}&=&T\int \frac{d^{2}k}{(2\pi )^{2}}\Delta _{Dk}^{BCS}\ \cos
k_{x}\sum_{n}e^{i\omega _{n}0^{+}}G({\bf k},i\omega _{n}) \nonumber \\
& &G(-{\bf k},-i\omega
_{n}),  \label{matsug}
\end{eqnarray}
can be cast in real frequency as 
\begin{eqnarray}
\varphi _{x} &=&-\int \frac{d^{2}k}{(2\pi )^{2}}\Delta _{Dk}^{BCS}\;\cos
k_{x}\nonumber \\
& &\int d\omega d\omega ^{\prime }\frac{A({\bf k},\omega )A(-{\bf k}%
,\omega ^{\prime })}{\omega +\omega ^{\prime }}\tanh (\frac{\omega }{2T}),
\label{8}
\end{eqnarray}
where $n_{F}(\omega )=1/(1+\exp ((\omega -\mu _{ef})/T))$. Eq. (\ref{8})
reduces to the linearized version of the usual BCS Eq. (\ref{5}) when $A(%
{\bf k},\omega )=\delta (\omega -{\bar{\epsilon}}_{k})$. As discussed in
previous works, assuming $\xi $ approximately constant or with a weak $T$-
dependence, two different regimes due to the AFSF can be distinguished as a
function of $T$ in the behavior of $A({\bf k},\omega )$ \cite{chu,vil}: (i)
For $T<<\omega _{sf}$, quantum contributions dominate the behavior of the
self-energy (\ref{7}) and $A({\bf k},\omega )$ exhibits Fermi liquid-like
quasiparticle peaks, even for Fermi points near ${\bf k}_{hs}$. Within this
regime, no relevant qualitative changes in comparison with the BCS
description are expected in the solution of the gap Eq. (\ref{8}). (ii) For $%
T>>\omega _{sf}$, and for ${\bf k}$- points satisfying $\xi >>\xi _{th}$,
classical effects (introduced by the $m=0$-Matsubara frequency in (\ref{7}))
dominate. The AFSF can be considered as quasistatic and $A({\bf k}%
_{hs},\omega )$ exhibits a shadow-band structure for large enough values of $%
g^{\prime }$ \cite{chu,vil}. This implies a transfer of spectral weight from
low to high frequencies, and we expect an effective blurring of the large
density of states near the VHS with a concomitant decrease of $T_{c}$.

The above qualitative issues are confirmed by our numerical calculations, as
illustrated in Figure 2. We restrict ourselves to the case with finite $%
t^{\prime }=-0.45t$, which reproduces the observed Fermi surface of YBCO and
B2212 and to the case of $t_{AB}=2t$, for which the optimal doping and the
maximum $T_{c}$ predicted by the BCS approximation are in good agreement
with experiments (for these parameters, the ratio $t^{\prime }/t_{eff}\sim
0.27$). To compute $T_{c}$ we evaluate $\mu _{eff}$ and $\tau $ from 
\begin{eqnarray}
n&=&1+T\int \frac{d^{2}k}{(2\pi )^{2}}\sum_{n}e^{i\omega _{n}0^{+}}\text{Re}G(%
{\bf k},i\omega _{n}),\nonumber \\
\tau &=& T\int \frac{d^{2}k}{(2\pi )^{2}}
\sum_{n}e^{i\omega _{n}0^{+}}\cos k_{x}\text{Re}G({\bf k},i\omega _{n})
\label{9}
\end{eqnarray}
\begin{figure}
\narrowtext
\epsfxsize=3.3truein
\vbox{\hskip 0.05truein \epsffile{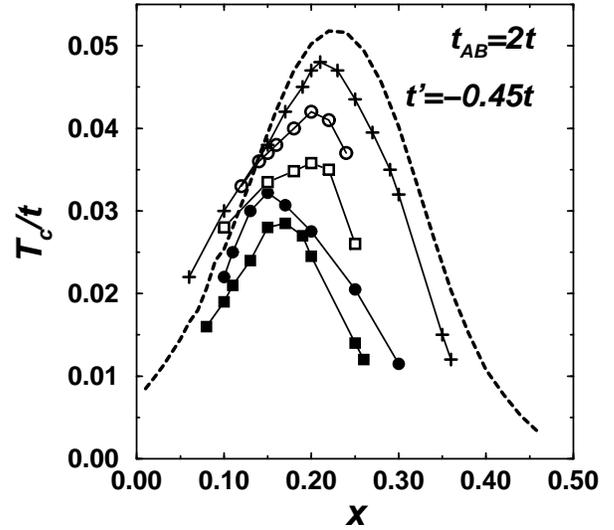}}
\medskip
\caption{ Critical temperatures $T_c$ in units of $t$ as a function of
the doping $x$, for $t_{AB}=2t$ and $t^{\prime}=-0.45t$, considering AFSF.
The dashed line (crosses) corresponds to the BCS-solution in the
thermodynamic limit (using $k$-discretization). Circles correspond to $%
g^{\prime}=3t$, and $\xi=2.5$ (open ones) and $\xi=10$ (filled ones).
Squares correspond to $g^{\prime}=10t$ and $\xi=2.5$ (open ones) and $\xi=4$
(filled ones).}
\end{figure}
and afterwards check if there exists (or not) a solution of the linearized
gap Eq. (\ref{matsug}). A cut-off $\Omega _{m}=432\omega _{sf}$ was chosen
for the $\nu _{m}$-Matsubara summation in (\ref{7}) while 
another one $\Omega _{n}=46t$, assuming $t=250meV$, 
was used for the $\omega _{n}$%
-Matsubara summation in (\ref{matsug}) and (\ref{9}). The sum over the tails
was approximated by an Euler-Mc Laurin formula. The $k$-integral in Eq. (\ref
{7}) is calculated with a relative precision of $10^{-6}$, while 
a fixed
finite mesh was used to evaluate the $k$-integrals in Eqs. (\ref{matsug})
and (\ref{9}). In the latter step, some precision is lost and the BCS- $%
T_{c} $ vs $x$, calculated in the thermodynamic limit, is not exactly
recovered as indicated in Fig. 2. We considered $\xi =2.5,\;(\omega
_{sf}=0.0256t)$ and $\xi =4,\;(\omega _{sf}=0.01t)$, which are supposed to
be representative of YBCO and LaSrCuO, respectively \cite{susc} near optimal
doping, and $\xi =10,\;(\omega _{sf}=0.0016t)$ which could be realistic only
for strongly underdoped materials. Taking $g^{\prime }=3t$ a weak effect on $%
T_{c}$ is observed. For $g^{\prime }=10t$, which is expected to be
representative of the cuprates \cite{mon,vil}, a decrease in $T_{c}$ is
observed as $\xi $ increases. For the case of $\xi =10,g^{\prime }=10t$ (not
shown in the figure) the maximum $T_{c}$ is below $0.012t$.

\section{Discussion}

We have shown that for high enough next-nearest-neighbor hopping 
$t^{\prime} $, $d$-wave superconductivity is stabilized for doping $x\sim 0.35$ at the
mean-field level in the correlated hopping model (\ref{1}). The
corresponding critical temperature $T_{c}$ has the right order of magnitude 
($\sim 70-100K$).
 Our calculations are in 2D and this result may be,
of course, affected by the fluctuations of the superconducting state,
which has not been taken into account in the present work.
However,
within the optimally doped to overdoped regime, the superfluid density
is large and the critical temperature corrected by these latter
fluctuations is expected to be close to the mean-field BCS one, even
when the transition is Kosterlitz-Thouless rather than BCS-like
\cite{em}. 
For dopings accessible experimentally
in the cuprates ($x<0.5$) the $s$-wave superconducting instability is
suppressed. Including the effect of antiferromagnetic spin fluctuations
(AFSF), we find that three different regimes remain as a function of doping
in the realistic range: (i) long-range antiferromagnetism near half filling,
(ii) $d$-wave superconductivity in a pure Fermi liquid (usual BCS) scenario
within the overdoped regime; (iii) $d$-wave superconductivity in the
presence of AFSF within the underdoped to optimally doped regime. The
maximum $T_{c}$ within our BCS treatment depends on the strength of the
interaction and the position of the VHS. The effect of the
spin-fluctuations, however, modifies the BCS picture within this regime. The
effective pairing interaction of (\ref{1}) has the form of an exchange
coupling, being ineffective in the Neel state \cite{afd}, but providing an
explicit coupling in the regime of doping for which the AF correlations are
short ranged. We considered particular values of the model parameters for
which optimal doping, $T_{c}$ and shape of the Fermi surface are in
agreement with experiments and further examined the effect of the
non-trivial temperature behavior induced by the existence of hot spots in
the Fermi surface in the presence of AFSF. We found that $T_{c}$ decreases
as the correlation length increases in correspondence with experiments.
For optimally doped materials, we expect that the magnitude of
$T_{c}$ is dominated by the effect of spin fluctuations
rather than by phase fluctuations  of the
superconducting order parameter.
Details of the coupling with AFSF, renormalizations of the bare
interactions, and eventual consequences upon the pseudogap behavior requires
a treatment of the full T-matrix of the effective pairing interaction on
equal footing with the spin susceptibility and is left for future studies.

\section*{Acknowledgments}

L. A. thanks the Max-Plank Institut f\"ur Physik komplexer Systeme, where
most of this work has been done, for its hospitality, T. Dahm for useful
discussions and J. Schmalian for useful comments. A. A. A. is partialy
supported by CONICET.


\begin{references}
\bibitem{an}  P. W. Anderson, Science {\bf 235}, 1196 (1987).

\bibitem{t-J1}  M. Ogata, M.U. Luchini, S. Sorella, and F.F. Assaad, Phys.
Rev. Lett. {\bf 66}, 2388 (1991); E. Dagotto and J. Riera, {\em ibid} {\bf 70%
}, 682 (1994); E. Heeb and T. M. Rice, Europhys. Lett. {\bf 27}, 673 (1994);
C.D. Batista, L.O. Manuel, H.A. Ceccatto, and A.A. Aligia{\em {\it ,} ibid} 
{\bf 38}, 147 (1997).

\bibitem{pla}  N. M. Plakida, V.S. Oudovenko, P. Horsch, and A.I.
Liechtenstein, Phys. Rev. B {\bf 55}, 11997 (1997).

\bibitem{zha}  F.C. Zhang and T.M. Rice, Phys. Rev. B {\bf 37}, 3759 (1988).

\bibitem{tpri}  C.D. Batista and A.A. Aligia, Physica C {\bf 264}, 319
(1996); references therein

\bibitem{fei}  L.F. Feiner, J.H. Jefferson, and R. Raimondi, Phys. Rev.
Lett. {\bf 76}, 4939 (1996); references therein.

\bibitem{bel}  V.I. Belinicher, A.L. Chernyshev, and V.A. Shubin, Phys. Rev.
B {\bf 54}, 14914 (1996); references therein.

\bibitem{fed}  H.B. Sch\"{u}ttler and A.J. Fedro Phys. Rev. B {\bf 45}, 7588
(1992);

\bibitem{sim}  M.E. Simon, M. Bali\~{n}a and A.A. Aligia, Physica C {\bf 206}%
, 297 (1993); M.E. Simon, A.A. Aligia and E. Gagliano, Phys. Rev. B {\bf 56}%
, 5637 (1997); references therein.

\bibitem{hub}  H. Lin, J. Hirsch and D. Scalapino, Phys. Rev B {\bf 37},
7359 (1988); A. Moreo, {\em ibid} {\bf 45}, 5059 (1992); E. Dagotto, Rev.
Mod. Phys. {\bf 66}, 763 (1994); S. Zhang, J. Carlson and J.E. Gubernatis,
Phys. Rev. Lett {\bf 78}, 4486 (1997).

\bibitem{arpse}  J.M. Harris, Z.X. Shen, P.J. White, D.S. Marshall, M.C.
Schabel, J.N. Eckstein, and I. Bozovic, Phys. Rev. B {\bf 54}, R15665
(1996); A. G. Loeser, Z.X. Shen, D.S. Dessau, D.S. Marshall, C.H. Park, P.
Fournier and A. Kapitulnik, Science {\bf 273}, 325 (1996); H. Ding, T.
Tokoya, J.C. Campuzano, T. Takahashi, M. Randeira, M.R. Norman, A.
Kapitulnik, T. Mochika, K. Kadowaki, and J. Giapintzakis, Nature {\bf 382},
51 (1996); G. Blumberg, M. Kang. and C. Kendziora, Science {\bf 278} 1427
(1997); M. R. Norman, H. Ding and K. Kadowaki, Phys. Rev. Lett. {\bf 79}
3506 (1997).

\bibitem{precaf}  A. P. Kampf, J. R. Schrieffer, Phys. Rev. B {\bf 42}, 7967
(1990); A. V. Chubukov and D. Morr, Phys. Rep. {\bf 288}, 355 (1997).

\bibitem{rande}  M. Randeria, cond-mat/9710223; Y. M. Vilk, S. Allen, H.
Touchette, S. Moukouri, L. Chen and A.-M. S. Tremblay, cond-mat/9710013.

\bibitem{eng}  J. Engelbrecht, A. Nazarenko and E. Dagotto, Phys. Rev. B 
{\bf 57} 13406 (1998).

\bibitem{em} V. J. Emery and S. A. Kivelson, Nature {\bf 374}, 434
(1995).

\bibitem{other} H. Fukuyama and H. Kohno, Physica C {\bf 282-287}, 
124 (1997); P. A.
Lee and X. G. Wen, Phys. Rev. Lett. {\bf 76}, 503 (1996).

\bibitem{negu}  A. Moreo and D. Scalapino, Phys. Rev. Lett {\bf 66}, 946
(1991); M. Randeria, N. Trivedi, A. Moreo, and R.T. Scalettar,{\em \ ibid} 
{\bf 69}, 2001 (1992); N. Trivedi and M. Randeria, Phys. Rev. Lett {\bf 75},
312 (1995); R. Haussmann, Phys. Rev. B {\bf 49}, 12975 (1994); R. Micnas, S.
Robaszkiewicz, and T. Kostyrko,{\em \ ibid} {\bf 52}, 16223 (1995).

\bibitem{feh}  R. Fehrenbacher and M. R. Norman, Phys. Rev. Lett. {\bf 74},
3884 (1995).

\bibitem{avh}  E. Dagotto, A. Nazarenko, and A. Moreo, Phys. Rev. Lett. {\bf %
74}, 310 (1995); A. Nazarenko, A. Moreo, E. Dagotto and J. Riera, Phys. Rev.
B {\bf 54}, R768 (1996).

\bibitem{com}  A.A. Aligia, F. Lema, M.E. Simon, and C.D. Batista, Phys.
Rev. Lett. {\bf 79}, 3793 (1997).

\bibitem{susc}  A. J. Millis, H. Monien and D. Pines, Phys. Rev. B {\bf 42},
167 (1990); V. Barzykin and D. Pines, Phys. Rev B {\bf 52}, 13585 (1995).

\bibitem{chu}  A. Chubukov and J. Schmalian, Phys. Rev. B {\bf 57} R11085
(1998); J. Schmalian, D. Pines and B. Stojkovich, Phys. Rev. Lett. {\bf 80}
3839 (1998).

\bibitem{tabmt}  L. Arrachea, A. Aligia and E. Gagliano, Phys. Rev. Lett. 
{\bf 76} 4396 (1996); references therein.

\bibitem{physc}  L. Arrachea and A. Aligia, Physica C {\bf 289}, 70 (1997);
references therein.

\bibitem{mon}  P. Montoux, A.V. Balatsky and D. Pines, Phys. Rev. Lett. {\bf %
67}, 3448 (1992); P. Montoux and D. Pines, Phys. Rev. Lett. {\bf 69}, 961
(1992); A. J. Millis, Phys. Rev. B {\bf 45}, 13047 (1992).

\bibitem{afd}  L. Arrachea and A. A. Aligia, Physica C (to be published).

\bibitem{tra}  For example, A.A. Aligia, J. Dorantes- D\'{a}vila, J.L.
Mor\'{a}n- L\'{o}pez and K.H. Bennemann, Phys. Rev. B {\bf 35}, 7053 (1987).

\bibitem{kik}  R. Kikuchi, Physica A {\bf 142}, 321 (1987).

\bibitem{vil}  Y. M. Vilk, Phys. Rev. B {\bf 55}, 3870 (1997); Y. M. Vilk
and A. M. S. Tremblay, Europhys. Lett. {\bf 33}, 159 (1996).
\end{references}
\end{document}